\documentclass[twocolumn,pre,showpacs,preprintnumbers,amsmath,amssymb,floatfix]{revtex4-1}
\usepackage{graphicx}
\usepackage{amsmath}
\usepackage{amssymb}
\usepackage{color}
\begin{document}

\title{Continuum percolation of polydisperse hyperspheres in infinite dimensions}

\author{Claudio Grimaldi}\affiliation{Laboratory of Physics of Complex Matter, Ecole Polytechnique F\'ed\'erale de
Lausanne, Station 3, CH-1015 Lausanne, Switzerland}
\date{}

\begin{abstract}
We analyze the critical connectivity of systems of penetrable $d$-dimensional spheres having size distributions 
in terms of weighed random geometrical graphs, in which vertex coordinates correspond to random
positions of the sphere centers and edges are formed between any two overlapping spheres. Edge weights
naturally arise from the different radii of two overlapping spheres. 
For the case in which the spheres have bounded size distributions, we show that clusters of connected spheres 
are tree-like for $d\rightarrow \infty$ and they contain no closed loops. In this case we find that 
the mean cluster size diverges at the percolation threshold density $\eta_c\rightarrow 2^{-d}$, independently of the
particular size distribution. We also show that the mean number of overlaps for a particle at criticality $z_c$ is smaller 
than unity, while $z_c\rightarrow 1$ only for spheres with fixed radii. We explain these features by showing that
in the large dimensionality limit the critical connectivity is dominated by the spheres with the largest size.
Assuming that closed loops can be neglected also for unbounded radii distributions, we find that the asymptotic
critical threshold for systems of spheres with radii following a lognormal distribution is no longer
universal, and that it can be smaller than $2^{-d}$ for $d\rightarrow\infty$.
\end{abstract}


\maketitle

\section{Introduction}
\label{intro}
Percolation phenomena are ubiquitous in many aspects of natural, technological, and social sciences, and they arise when
system-spanning clusters or components of, in some sense, connected objects form \cite{Stauffer1994,Sahimi2003}. 
A quantity of much interest is the percolation threshold, which marks the transition between the phase in which 
a giant component exists and the one in which it does not. In general, the percolation threshold is a nonuniversal
quantity, as it depends on the connectivity properties of the specific system under consideration \cite{TorquatoBook}. 
For example, in continuum percolation systems, where objects occupy positions in a continuous space, the threshold depends on
the shape of the objects \cite{Balberg1984,Berhan2007,Charlaix1986,Ambrosetti2010a,Mathew2012}, on their 
interactions \cite{Chiew1983,Miller2004,Wei2015}, as well as on the connectedness criteria \cite{Xu1988,Chiew1999}. 

In this article, we consider the infinite-dimensional limit of a paradigmatic example of continuum percolation: 
the Boolean-Poisson model \cite{Meester1996,Penrose2003}.
In this model, penetrable spheres with distributed radii have centers generated by a point Poisson process, and any two spheres
are considered connected if they overlap. For a given distribution of the radii, the percolation threshold is given by the
critical concentration $\eta_c$ of spheres, or by the critical volume fraction $\phi_c=1-e^{-\eta_c}$, such that a giant component 
of connected spheres first forms. Precise numerical estimates of 
$\eta_c$ have been obtained in two and three dimensions for, respectively, disks and spheres with fixed or distributed 
radii \cite{Lorenz2000,Quintanilla2001,Consiglio2003,Ogata2005,Quintanilla2007,Mertens2012,Torquato2012b,Sasidevan2013}.
The general trend observed by these investigations is that $\eta_c$ depends on the form of the distribution function
of the radii, and that it has its minimum when the sphere radii are monodisperse (i.e., when the spheres have identical size). 
This last point has been formally confirmed in Ref.~\cite{Meester1994}, although it may not hold true in the limit of infinite 
dimensions $d$ \cite{Gouere2013,Gouere2014}.
 
Here we show that for bounded distributions of the radii, that is for polydisperse spheres with a maximum finite value of the radius,
the percolation threshold of the Boolean-Poisson model tends asymptotically to a universal constant as $d\rightarrow \infty$, provided
that the radii distribution is independent of $d$.
This constant coincides with the value found in Refs.~\cite{Penrose1996,Torquato2012} for spheres of identical radii, $\eta_c\rightarrow 2^{-d}$ ,
and it is independent of the particular form of the size distribution function. We interpret the universality of $\eta_c$ as being due to the
statistical irrelevance of the spheres with smaller radii: the onset of percolation is established effectively only by the subset of spheres
with maximum radius. Furthermore, we show that the mean number of connected spheres per particle at percolation, $z_c$, is less than 
unity for polydisperse distributions of the radii, while $z_c\rightarrow 1$ only in the limit of identical radii. This finding is analogous to 
what simulations have shown for the case of continuum percolation in three dimensional space of spherocylinders with length 
polydispersity \cite{Nigro2013}.

These results rest on the observation that closed loops of connected spheres can be neglected in the limit of large dimensions, 
as we show explicitly for the case of bounded radii distributions. In the hypothesis that closed loops are irrelevant also for
spheres of unbounded size, we show that $\eta_c$ for $d\rightarrow\infty$ is not universal, as it depends on the parameters of the 
distribution, and that it can be smaller than the critical threshold of monodisperse spheres, in contrast to what is expected for 
finite dimensions \cite{Meester1994}.

\section{The model}
\label{model}

To construct the Boolean model, we consider $N$ points placed independently and uniformly 
at random in a $d$-dimensional volume $V$. Each point is the center of a sphere with the radius
drawn independently and randomly from a given probability distribution function $\rho(R)$. If we denote
$N_1$ the number of spheres of radius $R_1$, $N_2$ the number of spheres of radius $R_2$, and so on, 
we can write the following without loss of generality:
\begin{equation}
\label{rho}
\rho(R)=\sum_i x_i\delta(R-R_i),
\end{equation}
where $x_i=N_i/N$ with $i=1,2,\ldots$ is the fraction of spheres of radius $R_i$. 

\begin{figure}[t]
\begin{center}
\includegraphics[scale=0.6,clip=true]{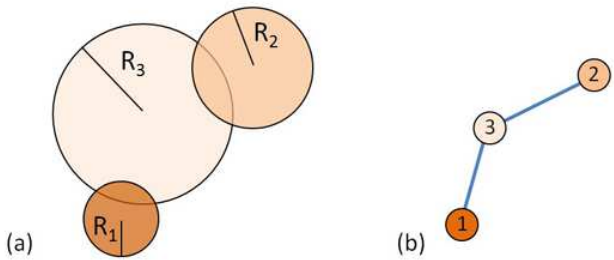}
\caption{(Color online) Connectedness criterion for spheres with different radii. (a) The spheres of radii $R_1$ and $R_2$
overlap the sphere of radius $R_3$, forming links between $R_3$ and $R_1$ and between $R_3$ and $R_2$. (b)
Corresponding cluster formed by nodes (sphere centers) labeled by the sphere radii and weighted links (solid lines)
connecting the nodes.}\label{fig1}
\end{center}
\end{figure}

Given any two spheres of radii, say, $R_i$ and $R_j$, we assign a link between their centers if the spheres
overlap, that is, if the distance $r$ between their center is smaller than $R_i+R_j$, as shown in Fig.~\ref{fig1}. 
We express this criterion for the formation of a link in terms of the connectedness function:
\begin{equation}
\label{connect}
f_{ij}(r)=\theta(R_i+R_j-r),
\end{equation}
where $\theta(x)=1$ for $x\geq 0$ and $\theta(x)=0$ for $x<0$ is the unit step function.

The set of sphere centers (nodes) and links (edges) forms a type of weighted random geometric graph, in which the probability
that an edge between two nodes is formed is weighted by the sphere radii. To see this, let us take a sphere of radius $R_i$
centered at the origin. The probability that a second sphere of radius $R_j$ forms a link with the first sphere is:
\begin{equation}
\label{vexij}
v_\textrm{ex}^{ij}=\frac{1}{V}\int\!d\mathbf{r}\,f_{ij}(r)=\Omega_d \frac{(R_i+R_j)^d}{V},
\end{equation} 
where $d\mathbf{r}$ is an infinitesimal $d$-dimensional volume element at the position $\mathbf{r}$ of the sphere of radius $R_j$, 
$\Omega_d=\pi^{d/2}/\Gamma(1+d/2)$ is the volume of a sphere of unit radius, and $\Gamma$ is the gamma function.
We note that $v_\textrm{ex}^{ij}$ defines also the excluded volume $V_\textrm{ex}^{ij}=\Omega_d (R_i+R_j)^d$ in 
units of $V$ between two spheres of different radii.

\section{Irrelevance of closed loops for $d\rightarrow\infty$}
\label{loops}

An important aspect of the topology of random geometric graphs is represented by closed loops (or cycles) of 
connected nodes. The most studied loop quantity is the three-nodes cycle $c^{(3)}_d$, often denoted the cluster coefficient, which 
gives the conditional probability that two nodes are connected given that both nodes are connected to a third one.
$c^{(3)}_d$ has been calculated for systems of spheres with identical radii and for any dimensionality \cite{Torquato2012,Dall2002}.  
The observation that $c^{(3)}_d$ vanishes exponentially as $d\rightarrow\infty$ indicates that random geometric graphs in large
dimensions have a locally tree-like structure. 
 
Using results from the theory of hard-sphere fluids, it is actually possible to show that, in the limit of large dimensions, closed 
loops are negligible also for any number of nodes and for bounded radii distributions. Random and weighted random geometric 
graphs have thus tree-like structures when $d\rightarrow \infty$. 
To see this, let us first consider the case of monodisperse spheres with radius $R_M$. We define an $n$-chain graph as a cluster 
of $n\geq 3$ nodes with $n-1$ edges such that every two consecutive edges, and
only those, have a common node. We denote as end-nodes the two nodes of an $n$-chain that each have only one edge.
The $n$-cycle coefficient $c^{(n)}_d$ is defined as the conditional probability that two nodes are connected given that they
are the end-nodes of an $n$-chain. Since the spheres have identical radii, we omit the subscripts in Eq.~\eqref{connect}, and we write 
the connectedness function as simply $f(r)=\theta(2R_M-r)$. 
From the definition of $c^{(n)}_d$, we can thus write:
\begin{equation}
\label{ncycle1}
c^{(n)}_d=\frac{\int dr^{(n)} f(\vert\mathbf{r}_1-\mathbf{r}_2\vert)
f(\vert\mathbf{r}_2-\mathbf{r}_3\vert)\cdots f(\vert\mathbf{r}_n-\mathbf{r}_1\vert)}
{\int dr^{(n)} f(\vert\mathbf{r}_1-\mathbf{r}_2\vert)
f(\vert\mathbf{r}_2-\mathbf{r}_3\vert)\cdots f(\vert\mathbf{r}_{n-1}-\mathbf{r}_n\vert)},
\end{equation}
where $dr^{(n)}=d\mathbf{r}_1d\mathbf{r}_2\cdots d\mathbf{r}_n$. 
Besides a prefactor, the above expression coincides with the cluster integral of a ring of $n$ hard-spheres
of radius $R_M$ \cite{Loeser1991}, as the Mayer function $f_\textrm{M}(r)$ for a fluid of hard-spheres is just 
$f_\textrm{M}(r)=-f(r)$ \cite{TorquatoBook,HansenMcDonald}.
To evaluate Eq.~\eqref{ncycle1} for $d\rightarrow\infty$, we thus use known results from the theory of hard-sphere
fluids in infinite dimensions. Noting that the denominator of Eq.~\eqref{ncycle1}
(i.e., the $n$-chain contribution) is simply $VV_\textrm{ex}^{n-1}$ \cite{Wyler1987}, where $V_\textrm{ex}=\Omega_d 2^dR_M^d$ is
the excluded volume for spheres of identical radius $R_M$, and introducing the Fourier transform
$\hat{f}(\mathbf{q})$ of the connectedness function we rewrite Eq.~\eqref{ncycle1} as:
\begin{equation}
\label{ncycle2}
c^{(n)}_d=\frac{1}{V_\textrm{ex}^{n-1}}\int\frac{d\mathbf{q}}{(2\pi)^d}\hat{f}(\mathbf{q})^n.
\end{equation}
The integration in Eq.~\eqref{ncycle2} for $d\rightarrow\infty$ has been worked out in Ref.~\cite{Loeser1991} 
(see also Ref.\cite{Frisch1999}), so that the $n$-cycle coefficient reduces to:
\begin{equation}
\label{ncycle3}
c^{(n)}_d\rightarrow \sqrt{\frac{n-2}{\pi d(n-1)}}\left(\frac{n}{n-2}\right)^{n/2}\left[\frac{n^{n-2}}{(n-1)^{n-1}}\right]^{d/2},
\end{equation}
from which we see that closed loops of any number $n$ of nodes are exponentially small as $d\rightarrow\infty$,
because the quantity within square brackets is less than unity for $n\geq 3$.

Let us now consider the $n$-cycle coefficient for the case of polydisperse spheres. Using Eq.~\eqref{connect}
for the connectedness function, we generalize Eq.~\eqref{ncycle1} as follows:
\begin{equation}
\label{ncycle4}
\langle c_d\rangle^{(n)}=\frac{\langle \mathcal{C}_{i_1,\ldots ,i_n}^{(n)}\rangle_{i_1,\ldots ,i_n}}
{\langle \mathcal{V}_{i_1,\ldots ,i_n}^{(n)}\rangle_{i_1,\ldots ,i_n}},
\end{equation}
where
\begin{align}
\label{ncycle5}
\mathcal{C}_{i_1,\ldots ,i_n}^{(n)}=\int dr^{(n)}& f_{i_1i_2}(\vert\mathbf{r}_1-\mathbf{r}_2\vert)
f_{i_2i_3}(\vert\mathbf{r}_2-\mathbf{r}_3\vert)\cdots \nonumber\\
&\cdots\times f_{i_ni_1}(\vert\mathbf{r}_n-\mathbf{r}_1\vert), \\
\label{Vexn}
\mathcal{V}^{(n)}_{i_1,\ldots ,i_n}=\int dr^{(n)}& f_{i_1i_2}(\vert\mathbf{r}_1-\mathbf{r}_2\vert)
f_{i_2i_3}(\vert\mathbf{r}_2-\mathbf{r}_3\vert)\cdots\nonumber\\
&\cdots\times f_{i_{n-1}i_n}(\vert\mathbf{r}_{n-1}-\mathbf{r}_n\vert),
\end{align}
and 
\begin{equation}
\label{ave1}
\langle(\cdots)\rangle_{i_1,\ldots ,i_n}=\sum_{i_1,\ldots,i_n}x_{i_1}x_{i_2}\cdots x_{i_n}(\cdots)
\end{equation}
denotes a multiple average over the radii $R_{i_1}, R_{i_2}, \ldots, R_{i_n}$. In the appendix, we show that for 
bounded distributions of radii, the $n$-cycle coefficient in the limit $d\rightarrow\infty$ is such that:
\begin{equation}
\label{ncycle6}
\langle c_d\rangle^{(n)}\leq c^{(n)}_d\chi_d^{(n)},
\end{equation} 
where $c^{(n)}_d$ is the $n$-cycle coefficient for identical radii, Eq.~\eqref{ncycle3}, and $\chi_d^{(n)}\propto d^a$,
where $a$ is a nonnegative constant. Since the exponential decay of $c^{(n)}$ for $d\rightarrow\infty$ is stronger
than the power-law increase of $\chi_d^{(n)}$, we see thus that
also for the case of polydisperse spheres for bounded radii distributions, the $n$-cycle coefficient vanishes for any $n\geq 3$. 

\section{Size of finite components}
\label{size}

The observation made in the previous section that closed loops are irrelevant in the large dimensional limit
of the Boolean model allows us to consider the components of the associated weighted random geometric graph as effectively
having a tree-like structure. This leads to a considerable simplification, as we can take the formalism
of the theory of random graphs (see, e.g., Refs.~\cite{Newman2001,Albert2002,Boccaletti2006}) and generalize it to 
the case in which nodes have weights.

\subsection{Multidegree distributions}
\label{multi}

We start by considering the multidegree distribution of a node of type $i$, defined as
the probability $p_i(1,k_1;2,k_2;\dots)$ that a sphere of radius $R_i$ is connected to $k_1$ spheres of 
radius $R_1$, $k_2$ spheres of radius $R_2$, and so on. Since the radii are randomly and independently 
distributed among the $N$ nodes, $p_i(1,k_1;2,k_2;\dots)$ is just a product of binomial 
distributions $p_{ij}(k_j)$ (with $j=1,2,\ldots$), each giving the probability that $k_j$ spheres of radius $R_j$ 
overlap the sphere of radius $R_i$:
\begin{equation}
\label{binomial1}
p_i(1,k_1;2,k_2;\dots)=\prod_j p_{ij}(k_j),
\end{equation}
with 
\begin{equation}
\label{binomial2}
p_{ij}(k_j)=\binom{N_j-\delta_{i,j}}{k_j}
(v_\textrm{ex}^{ij})^{k_j}(1-v_\textrm{ex}^{ij})^{N_j-\delta_{i,j}-k_j},
\end{equation}
where $N_j$ (with $j=1,2,\ldots$) is the number of spheres of radius $R_j$, $v_\textrm{ex}^{ij}$ are 
the overlap probabilities given in Eq.~\eqref{vexij}, and $\delta_{i,j}$ is the Kronecker symbol.

We next consider for all $i$ the limit $N_i\rightarrow\infty$ such that $N_i/V=x_i\rho$ remains finite, where $\rho=N/V$ is the
total number density. In this limit, Eq.~(\ref{binomial2}) reduces to a Poisson distribution:
\begin{equation}
\label{poisson1}
p_{ij}(k_j)=\frac{z_{ij}^{k_j}}{k_j!}e^{-z_{ij}},
\end{equation}
where
\begin{equation}
\label{zij}
z_{ij}=\sum_k kp_{ij}(k)=x_j\rho\Omega_d(R_i+R_j)^d
\end{equation}
is the average number of spheres with radius $R_j$ that overlap a given sphere of radius $R_i$. 

In addition to the node degree distribution $p_i(1,k_1;2,k_2;\dots)$, for the following analysis we will 
also need the excess node degree distribution $q_{ji}(1,k_1;2,k_2;\ldots)$, defined as the conditional probability that a
sphere of radius $R_j$ is connected to $k_l$ spheres of radius $R_l$ (with $l=1,2,\ldots$), given that
it is connected to a sphere of radius $R_i$.
This task is simplified by the irrelevance of closed loops in the large dimensionality limit. 
In this case, indeed, if we select at random an edge connecting a node of type $j$ with a node of type $i$, 
the $j$ node attached to the edge is $k_i$ times more likely to have degree $k_i$
than degree $1$ with nodes of type $i$. Its degree distribution will thus be proportional to 
$k_i p_j(1,k_1;2,k_2;\dots)$.
The excess degree distribution of a $j$ node that has $k_i$ edges with nodes of type $i$ other than the edge
with the node $i$ to which is attached is 
thus \cite{Leicht2009}:
\begin{equation}
\label{excess1}
q_{ji}(1,k_1;2,k_2,\ldots)=\frac{(k_i\!+\!1)p_j(1,k_1;\ldots;i,k_i\!+\!1;\ldots)}{\sum_\mathbf{k}(k_i\!+\!1)p_j(1,k_1;\ldots;i,k_i\!+\!1;\ldots)},
\end{equation}
where $\sum_\mathbf{k}=\sum_{k_1,k_2,\ldots}$. From Eqs.~\eqref{binomial1} and \eqref{poisson1}, $q_{ji}(1,k_1;2,k_2,\ldots)$
reduces simply to:
\begin{align}
\label{excess2}
q_{ji}(1,k_1;2,k_2,\ldots)&=\frac{(k_i+1)p_{ji}(k_i+1)}{z_{ji}}\prod_{l\neq i}p_{jl}(k_l)\nonumber \\
&=\prod_l p_{jl}(k_l),
\end{align}
where we have used $(k_i+1)p_{ji}(k_i+1)=z_{ji}p_{ji}(k_i)$. Equation \eqref{excess2} states thus the well-known
property that the excess degree distribution coincides with the node degree distribution when this is Poissonian \cite{Newman2001}.

\subsection{Mean cluster size in the subcritical regime}
\label{size}

We exploit now the statistical irrelevance of closed loops discussed in Sec.~\ref{loops} to
find the mean size $S$ of finite clusters of connected spheres as $d\rightarrow\infty$. In doing so,
we shall first keep the form of the degree distributions unspecified, and apply Eqs.~\eqref{poisson1} and \eqref{excess2}
only at the end of the calculation.

\begin{figure}[t]
\begin{center}
\includegraphics[scale=0.6,clip=true]{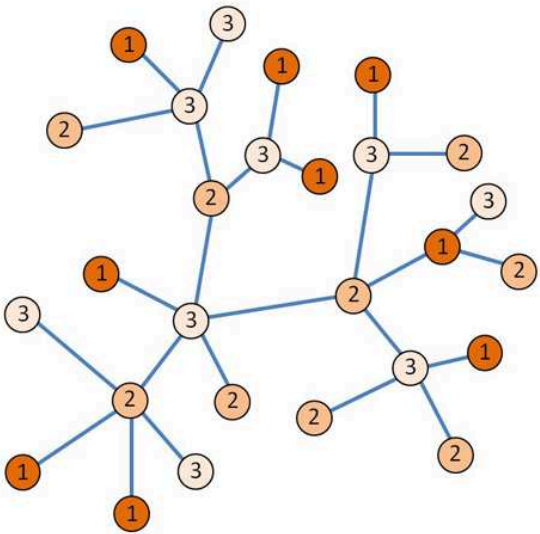}
\caption{(Color online) Schematic representation of a finite tree-like cluster formed by connected spheres
of radii $R_1$, $R_2$, and $R_3$. Each node label corresponds to the value of the radius of the sphere 
attached to the node.}\label{fig2}
\end{center}
\end{figure}

Let us start
by considering a randomly selected node that has probability $x_i$ of being 
occupied by a sphere of radius $R_i$. Due to the general tree-like structure of the graph, the cluster to which 
the selected node belongs is formed by branches attached to the node according to the degree 
distribution $p_i(1,k_1;2,k_2;\ldots)$, as schematically shown in Fig.~\ref{fig2}.
The mean size $S_i$ of the cluster to which the selected node belongs is thus:
\begin{equation}
\label{S1}
S_i=x_i+x_i\sum_\mathbf{k}p_i(1,k_1;2,k_2;\ldots)\sum_j k_jT_{ij},
\end{equation} 
where $T_{ij}$ is the mean cluster size of one of the $k_j$ branches attached to the selected node.  
Since the clusters have a tree-like structure, $T_{ij}$ is given by the mass (unity) of one neighbor of the selected node,
plus the mean cluster size of each of the remaining subbranches attached to the neighbor. 
To find $T_{ij}$, we thus need the excess degree distribution $q_{ji}(1,k_1;2,k_2;\ldots)$ of a sphere of radius $R_j$ 
connected to the selected node of type $i$:
\begin{equation}
\label{T1}
T_{ij}=1+\sum_\mathbf{k}q_{ji}(1,k_1;2,k_2;\ldots)\sum_l k_l T_{jl}.
\end{equation}

Equations \eqref{S1} and \eqref{T1} are quite general, as they apply also to tree-like graphs with degree
distributions that are not reducible to a multiplication of Poissonian probabilitites. Interestingly, similar equations are
found in the calculation of finite size components of multigraphs (also denoted multiplex networks), formed by different
networks, each having particular node properties, coupled together \cite{Leicht2009,Allard2009}. The Boolean-Poisson model 
with random radii can thus be viewed also as a particular type of multigraph, in which each individual network is constituted by nodes
occupied by spheres of a given radius.

Let us now use the results of Sec.~\ref{multi} and rewrite Eqs.~\eqref{S1} and \eqref{T1} by substituting
$p_i(1,k_1;2,k_2;\ldots)$ and $q_{ji}(1,k_1;2,k_2;\ldots)$ with, respectively, Eqs.~\eqref{binomial1} and \eqref{excess2}:
\begin{align}
\label{S2}
S_i&=x_i+x_i\sum_j\sum_k k p_{ij}(k) T_j=x_i+x_i\sum_j z_{ij}T_j, \\
\label{T2}
T_j&=1+\sum_l\sum_k kp_{jl}(k)T_l=1+\sum_l z_{jl}T_l,
\end{align}
where we have used Eq.~\eqref{zij} and the fact that $T_{ij}$ depends only on the neighbor ($j$) of the selected 
node, i.e., $T_{ij}=T_j$.

The mean cluster size is given by $S=\sum_i S_i$, which from Eq.~\eqref{S2} reduces to: $S=1+\sum_{ij}x_iz_{ij}T_j$. 
This relation is obtained also if we multiply both sides of Eq.~\eqref{T2} by $x_j$ and sum over $j$. We can thus write:
\begin{equation}
\label{S3}
S=\sum_j x_j T_j,
\end{equation}
which states that $S$ is just the average over the sphere radii of the mean cluster size of the branches.

\subsection{Equivalence with the Ornstein-Zernike equation for the pair-connectedness}
\label{OZ}

In continuum percolation theory, cluster statistics are often studied using
the formalism of pair-connectedness correlation functions \cite{TorquatoBook,Coniglio1977,Stell1996}, 
which exploits well developed techniques of liquid state theory. This method has been recently
used to studying percolation of monodisperse spheres in large dimensions \cite{Torquato2012}.

As long as closed loops can be neglected, the network formalism discussed above
and the pair-connectedness functions method give identical results, provided that the second-virial 
approximation is taken. To see how this equivalence holds
true for the Boolean model in large dimensions, let us first consider the pair-connectedness
function $P_{ij}(\mathbf{r}-\mathbf{r}')$, defined such that $x_ix_j\rho^2P_{ij}(\mathbf{r}-\mathbf{r}')d\mathbf{r}d\mathbf{r}'$ 
is the probability of finding two spheres of radii $R_i$ and $R_j$ within the volume elements $d\mathbf{r}$ and $d\mathbf{r}'$
centered respectively in $\mathbf{r}$ and $\mathbf {r}'$, given that they belong to the same cluster.
The mean cluster size $S$ is given in terms of $P_{ij}(\mathbf{r}-\mathbf{r}')$ by the following relation\cite{Otten2011}:
\begin{equation}
\label{OZ1}
S=1+\rho\sum_{i,j}x_i x_jP_{ij},
\end{equation}
where $P_{ij}=\int d\mathbf{r}P_{ij}(\mathbf{r})$. $P_{ij}$ is the solution
of the pair connectedness analog of the Ornstein-Zernike equation of the liquid
state theory of fluids:
\begin{equation}
\label{OZ2}
P_{ij}=C_{ij}+\rho\sum_l x_l C_{il}P_{lj},
\end{equation}
where $C_{ij}=\int d\mathbf{r}C_{ij}(\mathbf{r})$ is the volume integral of the
direct pair connectedness function $C_{ij}(\mathbf{r})$, which describes
short-range connectivity correlations. Let us introduce the
quantity $\widetilde{T}_i$ defined as:
\begin{equation}
\label{OZ3}
\widetilde{T}_i=1+\rho\sum_j x_j P_{ij}.
\end{equation}
The use of the above expression reduces Eq.~\eqref{OZ1} to:
\begin{align}
\label{OZ4a}
S&=\sum_i x_i+\rho\sum_{i,j}x_i x_jP_{ij}=\sum_i x_i\left(1+\rho\sum_j x_j P_{ij}\right)\nonumber \\
&=\sum_i x_i\widetilde{T}_i,
\end{align}
while inserting Eq.\eqref{OZ2} into Eq.~\eqref{OZ3} leads to:
\begin{align}
\label{OZ4b}
\widetilde{T}_i&=1+\rho\sum_j x_j \left(C_{ij}+\rho\sum_l x_l C_{il}P_{lj}\right)\nonumber \\
&=1+\rho\sum_j x_j C_{ij}+\rho\sum_j x_j C_{ij}(\widetilde{T}_j-1)\nonumber \\
&=1+\rho\sum_j x_j C_{ij}\widetilde{T}_j.
\end{align}
We see that Eqs.~\eqref{OZ4a} and \eqref{OZ4b} are identical to respectively
Eqs.~\eqref{S3} and \eqref{T2} if we identify $\rho x_j C_{ij}$ with $z_{ij}$. From
Eq.~\eqref{zij}, we obtain thus:
\begin{equation}
\label{C2}
C_{ij}=\frac{z_{ij}}{\rho x_j}=\Omega_d (R_i+R_j)^d,
\end{equation}
which corresponds to take the volume integral of the second-virial approximation $C_{ij}(\mathbf{r})=C_{ij}^{(2)}(\mathbf{r})=f_{ij}(r)$
for the direct pair-connectedness function.
This is not surprising, because in the density expansion of the direct pair-connectedness function, 
$C_{ij}(\mathbf{r})=\sum_{n\geq 2} \rho^{n-2}C_{ij}^{(n)}(\mathbf{r})$, the terms with $n\geq 3$ contain
at least one closed loop. 

\section{Universality of the percolation threshold}
\label{universal}

We proceed to find the percolation threshold for the Boolean-Poisson model of polydisperse
spheres in the large dimensionality limit. We shall consider the case of bounded distributions of the radii, 
for which we have shown in Sec.~\ref{loops} that closed loops of connected particles can be neglected 
for $d\rightarrow\infty$, and Eqs.~\eqref{T2} and \eqref{S3} are valid. To measure the sphere concentration 
we introduce the dimensionless density
\begin{equation}
\label{eta}
\eta=\rho\Omega_d\langle R^d\rangle_R=\rho\Omega_d\sum_i x_i R_i^d.
\end{equation}
The percolation threshold $\eta_c$ is defined as the smallest value of $\eta$ such that $S$ diverges. This definition is 
equivalent to finding the smallest pole of Eq.~\eqref{T2}, if it exists.

\subsection{Discrete radii distributions}
\label{discrete}
We first consider the case in which the spheres have a finite number $M$ of radii:
\begin{equation}
\label{fin1}
\rho(R)=\sum_{i=1}^Mx_i\delta(R-R_i),
\end{equation}
so that using Eqs.~\eqref{zij}, \eqref{T2} and \eqref{S3} we rewrite the equations for the mean cluster size as:
\begin{align}
\label{S4}
S&=\sum_{i=1}^M x_iT_i,\\
\label{T4}
T_i&=1+\rho\Omega_d\sum_{j=1}^Mx_j(R_i+R_j)^d T_j.
\end{align}
Without loss of generality, we assume that $R_M$ is strictly the largest radius out of the $M$ possible values
of the radii, and we introduce $q_i=R_i/R_M$, which takes values smaller than the unity for all $i\neq M$.
For large $d$, the dimensionless density $\eta$ reduces to:
\begin{align}
\label{eta1}
\eta&=\rho\Omega_d\sum_{i=1}^M x_iR_i^d=\rho\Omega_dR_M^d\left[x_M+\sum_{i=1}^{M-1} x_i q_i^d\right]\nonumber \\
&\rightarrow\rho\Omega_dR_M^dx_M,
\end{align}
because $q_i^d$ goes exponentially to zero as $d\rightarrow \infty$ when $i\neq M$, and Eq.~\eqref{T4} becomes:
\begin{equation}
\label{T5}
T_i=1+2^d\eta\frac{1}{x_M}\sum_j x_j\left(\frac{q_i+q_j}{2}\right)^dT_j.
\end{equation}
We note that $\left(\frac{q_i+q_j}{2}\right)^d$ is vanishingly small as $d\rightarrow \infty$ unless $i=j=M$, 
for which it takes the value $1$. The smallest pole of Eq.~\eqref{T5} for large $d$ is thus the solution of:
\begin{equation}
\label{T6}
T_i=1+2^d\eta T_M\delta_{i,M},
\end{equation}
where $\delta_{i,j}$ is the Kronecker delta.  Equation~\eqref{T6}  is solved by $T_M=1/(1-2^d\eta)$ and
$T_i=1$ for $i\neq M$, so that the mean cluster size \eqref{S4} becomes:
\begin{equation}
\label{S5}
S=\sum_{i=1}^{M-1}x_i+x_M T_M=\frac{x_M}{1-2^d\eta},
\end{equation}
which diverges when
\begin{equation}
\label{etac1}
\eta\rightarrow\eta_c=\frac{1}{2^d}.
\end{equation}
The above expression for $\eta_c$ holds true for any sequence of occupation fractions $x_i$, independent of dimensionality, 
provided that $x_M\neq 0$. In particular, Eq.~\eqref{etac1} confirms and extends to $M>2$ the finding of a previous report that
spheres with two different radii have a universal critical threshold in infinite dimensions \cite{Gouere2014}. Note that $\eta_c=1/2^d$ is also the limit 
for infinite dimensions of the percolation threshold of monodisperse spheres with radius $R_M$, whose mean cluster size is 
given by Eq.~\eqref{S5} with $x_M=1$.

\begin{figure}[t]
\begin{center}
\includegraphics[scale=0.57,clip=true]{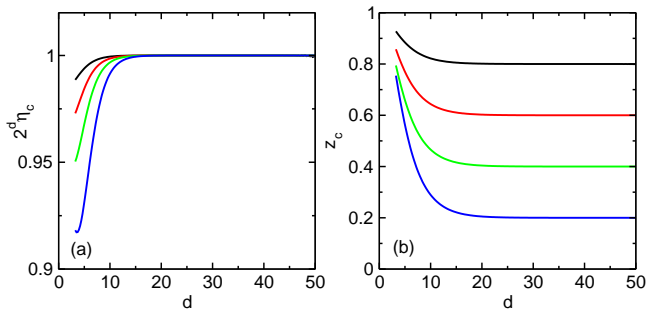}
\caption{(Color online) (a) Percolation threshold $\eta_c$ in units of the asymptotic value $2^{-d}$ as a function
of dimensionality for a discrete distribution of radii with $M=2$ and $R_1/R_2=1/2$. In this case Eq.~\eqref{T4} 
is a system of two linear equations so that $\eta_c$ can be calculated exactly for any $d$. $x_2=0.8$, 0.6$, 0.4$, and $0.2$
from the uppermost to the lowermost curves. (b) Corresponding values of the critical coordination number $z_c$. As
$d\rightarrow \infty$, $z_c$ tends asymptotically to $x_2$.}\label{fig3}
\end{center}
\end{figure}

The origin of the universality of $\eta_c$ can be traced back to the divergence of $T_M$, which indicates that the onset 
of a giant component of connected polydisperse spheres is established only by the subset of spheres with the maximum 
radius when $d\rightarrow\infty$. In other words, at $d\rightarrow\infty$ the contribution
to percolation of the smaller spheres vanishes, and the resulting $\eta_c$ is the critical threshold for a system of
monodisperse spheres of radius $R_M$. Following the observation that systems of polydisperse spheres with
$M$ different radii can be interpreted as a multinetwork of coupled $M$ subnetworks (see Sec.~\ref{size}), we see
that Eq.~\eqref{T6} is equivalent to decoupling the subnetworks associated with each radius, and that long-range connectivity
arises only from the network formed by spheres with radius $R_M$.

One interesting consequence of the irrelevance of smaller radii at percolation is that the critical average connectivity per particle,
\begin{equation}
\label{zc1}
z_c=\sum_{i,j}x_i x_j\rho_c\Omega_d(R_i+R_j)^d,
\end{equation}
reduces for $d\rightarrow\infty$ to:
\begin{equation}
\label{zc2}
z_c=2^d\rho_c\Omega_d R_M^d\sum_{i,j}x_i x_j\left(\frac{q_i+g_j}{2}\right)^d
\rightarrow2^d\eta_c x_M=x_M,
\end{equation}
where $\rho_c$ is the critical number density. For $x_M< 1$, the critical average connectivity is thus less than
the unity for $d\rightarrow \infty$, which must be contrasted to
$z_c\geq 1$ for systems constituting only of monodisperse spheres in any dimension \cite{Torquato2012}. 

For the binary case ($M=2$), Eq.~\eqref{T4} reduces to a system of two linear equations that can be solved exactly for any $d$. 
The resulting $\eta_c$ and $z_c$ are shown in Figs.~\ref{fig3}(a) and \ref{fig3}(b), respectively, for $R_2=2R_1$ and
different values of the fraction $x_2$ of spheres of radius $R_2$. The asymptotic limits $\eta_c=2^{-d}$ and $z_c=x_2$ 
are recovered for sufficiently large values of $d$. 

\subsection{Continuous radii distributions}
\label{continuous}

Let us now consider the case in which the radii distribution $\rho(R)$ is a continuous bounded function
independent of $d$. We again denote by $R_M<\infty$ the maximum allowed radius, so that $\rho(R)=0$ for 
$R>R_M$, and we rewrite the equations for the mean cluster size in terms of continuous variables of the radii:
\begin{align}
\label{CS1}
S&=\langle T(R)\rangle_R, \\
\label{CT1}
T(R)&=1+\rho\Omega_d\langle (R+R')^dT(R')\rangle_{R'},
\end{align}
where $\langle(\cdots)\rangle_R=\int_0^{R_M}dR\rho(R)(\cdots)$. We expand the binomial power $(R+R')^d$
and use $\eta=\rho\Omega_d\langle R^d\rangle_R$ to write:
\begin{equation}
\label{CT2}
T(R)=1+\eta\sum_{k=0}^d\binom{d}{k}\frac{R^{d-k}}{\langle R^d\rangle_R}\langle R^kT(R)\rangle_R.
\end{equation}
If we multiply both sides of Eq.~\eqref{CT2} by $R^n/\langle R^n\rangle_R$, with $n=1$, $2$, $\ldots$, $d$, and average over $R$ 
we arrive at:
\begin{equation}
\label{CT3}
t(n)=1+\eta\sum_{k=0}^d\binom{d}{k}\frac{\langle R^{n+d-k}\rangle_R\langle R^k\rangle_R}{\langle R^d\rangle_R\langle R^n\rangle_R}t(k),
\end{equation}
where
\begin{equation}
\label{tn}
t(n)=\frac{\langle R^n T(R)\rangle_R}{\langle R^n\rangle_R}.
\end{equation}
From Eqs.~\eqref{CS1} and \eqref{tn} we see that the mean cluster size can be obtained from $S=t(0)$.

To solve Eq.~\eqref{CT3}, we note that for large $d$ the  binomial coefficient is strongly peaked at $k=d/2$ 
and takes the asymptotic form
\begin{equation}
\label{binomial}
\binom{d}{k}\simeq 2^d\sqrt{\frac{2}{\pi d}}e^{-\frac{2}{d}(k-d/2)^2}=\frac{2^{d+1}}{d}g\!\left(\frac{2k}{d}-1,\frac{1}{\sqrt{d}}\right),
\end{equation}
where $g(x,\sigma)=\exp(-x^2/2\sigma^2)/\sqrt{2\pi\sigma^2}$ is the Gaussian function. Provided that the radii distribution 
is bounded, the binomial coefficent dominates the $k$-dependence of the kernel.  
To see this, let us consider the $m$-th moment $\langle R^m\rangle_R=R_M^m\int_0^1 dy\rho(y) y^m$, where
$y=R/R_M$. For large $m$ the main contribution to the integral comes from $y$ close to $1$. Thus we make the quite general 
assumption that for $y\rightarrow 1$, the radii distribution behaves as $\rho(y)\propto (1-y)^{\alpha-1}$, with $\alpha>0$. 
Setting $t=m(1-y)$ for large $m$, we find
\begin{align}
\label{f2}
\langle R^m\rangle_R&\propto \frac{R_M^m}{m^\alpha}\int_0^m\!dt\, t^{\alpha-1}(1-t/m)^m\nonumber \\
&\simeq\frac{R_M^m}{m^\alpha} \int_0^\infty\!dt\, t^{\alpha-1} e^{-t}=\frac{R_M^m}{m^\alpha}\Gamma(\alpha),
\end{align}
so that for large $k$ the term $\langle R^{n+d-k}\rangle_R\langle R^k\rangle_R$ in Eq.~\eqref{CT3} is proportional
to $R_M^{n+d}/[(n+d-k)k]^\alpha$, which has a much weaker $k$-dependence than Eq.~\eqref{binomial}.
Next, we introduce $s=2n/d$ and $s'=2k/d$, which we treat as continuous variables for $d\rightarrow\infty$, and we replace the sum over $k$
 by an integral over $s'$: $\sum_{k=0}^d\rightarrow \frac{d}{2}\int_0^2 ds'$. If we denote $\tilde{t}(s)=t(ds/2)$ and $\tilde{t}(s')=t(ds'/2)$,
 Eq.~\eqref{CT3} becomes:
 \begin{align}
 \label{CT4}
 \tilde{t}(s)=1+2^d\eta\int_0^2 ds'&\left[g\!\left(s'-1,\frac{1}{\sqrt{d}}\right)\right.\nonumber\\
 &\left.\times\frac{\langle R^{d[1+(s-s')/2]}\rangle_R\langle R^{ds'/2}\rangle_R}{\langle R^d\rangle_R\langle R^{ds/2}\rangle_R}\tilde{t}(s')\right].
 \end{align} 
Since $g(s'-1,1/\sqrt{d})\rightarrow\delta(s'-1)$ for $d\rightarrow\infty$, the above expression reduces to: 
\begin{equation}
\label{CT5}
\tilde{t}(s)=1+\eta2^d \frac{\langle R^{d(1+s)/2}\rangle_R\langle R^{d/2}\rangle_R}{\langle R^d\rangle_R\langle R^{ds/2}\rangle_R}\tilde{t}(1),
\end{equation} 
from which we obtain the mean cluster size:
\begin{equation}
\label{CT6}
S=\tilde{t}(0)=1+\eta2^d\frac{\langle R^{d/2}\rangle_R^2}{\langle R^d\rangle_R}\tilde{t}(1).
\end{equation}
Setting $s=1$ in Eq.~\eqref{CT5}, we find $\tilde{t}(1)=(1-2^d\eta)^{-1}$, so that we arrive finally at:
\begin{equation}
\label{CT7}
S=\frac{1}{1-2^d\eta}\frac{\langle R^{d/2}\rangle_R^2}{\langle R^d\rangle_R},
\end{equation}
which, as found for the case of discrete distributions, diverges at 
\begin{equation}
\label{perc22}
\eta\rightarrow \eta_c=\frac{1}{2^d},
\end{equation}
independently of the particular form of the bounded distribution $\rho(R)$.

Using Eq.~\eqref{binomial} and considering the weak dependence of the moments of $R$,
we readily obtain the large dimensional limit of the critical average connectivity per particle: 
\begin{align}
\label{zc2}
z_c&=\rho_c\Omega_d\langle (R+R')^d\rangle_{R,R'}=\eta_c\sum_{k=0}^d\binom{d}{k}
\frac{\langle R^k\rangle_R \langle R^{d-k}\rangle_R}{\langle R^d\rangle_R}\nonumber\\
&\rightarrow\frac{\langle R^{d/2}\rangle_R^2}{\langle R^d\rangle_R},
\end{align}
from which we see that $z_c\leq 1$ for any bounded distribution of the radii. 
Note that from Eq.~\eqref{zc2} we recover $z_c=x_M$ when $\rho(R)$ is given by Eq.~\eqref{fin1}.

\begin{figure}[t]
\begin{center}
\includegraphics[scale=0.57,clip=true]{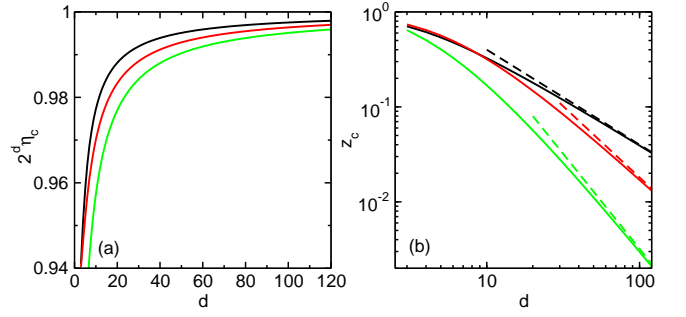}
\caption{(Color online) (a) Dimensional dependence of the percolation threshold $\eta_c$ in units of the asymptotic 
value $2^{-d}$ obtained from a numerical solution of Eq.~\eqref{CT3} for rectangular (upper curve), semicircular (middle curve),
and triangular (lower curve) distributions of the sphere radii. (b) Corresponding critical average connectivity per particle 
$z_c$ (solid curves). Dashed lines are the asymptotic results for $d\gg 1$: $z_c=4/d$ (rectangular distribution), 
$z_c=32/(\sqrt{\pi}d^{3/2})$ (semicircular distribution), and $z_c=32/d^2$ (triangular distribution).
 }\label{fig4}
\end{center}
\end{figure}

We complete this section by showing how the percolation threshold obtained from Eqs.~\eqref{CS1} and \eqref{CT1} evolves towards 
the asymptotic value $\eta_c=2^{-d}$ as $d$ increases. Toward that end, we consider radii distributions of rectangular, semicircular, and
triangular shapes, given respectively by $\rho(R)=1/R_M$, $\rho(R)=4\sqrt{(R_M/2)^2+(R-R_M/2)^2}/\pi$, and $\rho(R)=2(R_M-R)/R_M^2$,
for $R\leq R_M$ and zero otherwise. We calculate $\eta_c$ from the smallest pole of $S=t(0)$ obtained from a numerical solution of Eq.~\eqref{CT3}.
The resulting thresholds are very close to $2^{-d}$ for all $d$ considered, and they approach the asymptotic limit from below, as shown in Fig.~\ref{fig4}(a).
For the same cases of Fig.~\ref{fig4}(a), we have calculated also the $d$-dependence of $z_c$, shown in Fig.~\ref{fig4}(b) by solid lines, which
we compare with the asymptotic limits (dashed lines)  $z_c=4/d$, $z_c=32/(\sqrt{\pi}d^{3/2})$, and $z_c=32/d^2$ obtained from Eq.~\eqref{zc2} for
rectangular, semicircular, and triangular distributions of the radii, respectively.

\section{The case of unbounded distribution of the radii}
\label{unbounded}

Having established that $\eta_c$ is universal as $d\rightarrow\infty$ for bounded (and independent of $d$) distributions of the 
radii, it is natural to ask if universality holds true also when $\rho(R)$ is unbounded. Although we have shown the
irrelevance of closed loops limited to the case of bonded distributions, we shall nevertheless assume that $n$-cycle
coefficients are negligible also for unbounded $\rho(R)$, and that graph components have a tree-like structure. 
Let us consider the specific case of a lognormal distribution function:
\begin{equation}
\label{log1}
\rho(R)=\frac{1}{\sqrt{2\pi}\sigma R}\exp\!\left[-\frac{\ln^2(R/R_0)}{2\sigma^2}\right],
\end{equation}
where $R\in[0,\infty)$, $R_0$ is the median radius, and $\sigma$ is the standard deviation of $\ln(R)$. 
Equation \eqref{log1} is an interesting case-study, as the resulting $\eta_c$ and $z_c$ for asymptotically large 
$d$ can be found analytically. Using the $k$-th moment
$\langle R^k\rangle_R=R_0^k\exp(\sigma^2k^2/2)$, Eq.~\eqref{CT3} becomes:
\begin{equation}
\label{log2}
t(n)=1+\eta\sum_{k=0}^d\binom{d}{k}e^{\frac{1}{2}\sigma^2[(n+d-k)^2+k^2-n^2-d^2]}t(k),
\end{equation}
from which we express the mean cluster size as:
\begin{equation}
\label{log3}
S=t(0)=1+\eta\sum_{k=0}^d\binom{d}{k}e^{\sigma^2k(k-d)}t(k).
\end{equation}
For sufficiently large $d$, the only nonvanishing terms of the summation are those with $k=0$ and $k=d$, so that:
\begin{equation}
\label{log4}
S=1+\eta[S+t(d)],
\end{equation}
where from Eq.~\eqref{log2} $t(d)$ is given by:
\begin{equation}
\label{log5}
t(d)=1+\eta\sum_{k=0}^d\binom{d}{k}e^{\sigma^2(d-k)^2}t(k).
\end{equation}
For $d\rightarrow\infty$, $t(d)$ tends asymptotically to $t(d)= 1+\eta e^{\sigma^2d^2}t(0)$, as the term with $k=0$ 
dominates the sum over $k$ in Eq.~\eqref{log5}. We thus find that the mean cluster size, Eq.~\eqref{log4}, reduces to:
\begin{equation}
\label{log6}
S=\frac{1+\eta}{1-\eta-\eta^2e^{\sigma^2d^2}},
\end{equation}
which diverges at the asymptotical critical value,
\begin{equation}
\label{log7}
\eta\rightarrow\eta_c=e^{-\frac{1}{2}\sigma^2d^2}.
\end{equation}
The corresponding critical coordination number is
\begin{align}
\label{log8}
z_c&=\eta_c\sum_{k=0}^d\binom{d}{k}\frac{\langle R^k\rangle_R \langle R^{d-k}\rangle_R}{\langle R^d\rangle_R}
=\eta_c\sum_{k=0}^d\binom{d}{k}e^{\sigma^2k(k-d)}\nonumber\\
&\rightarrow 2\eta_c,
\end{align}
where we have again used the fact that for large $d$ only the terms $k=0$ and $k=d$ contribute to the summation.

\begin{figure}[t]
\begin{center}
\includegraphics[scale=0.57,clip=true]{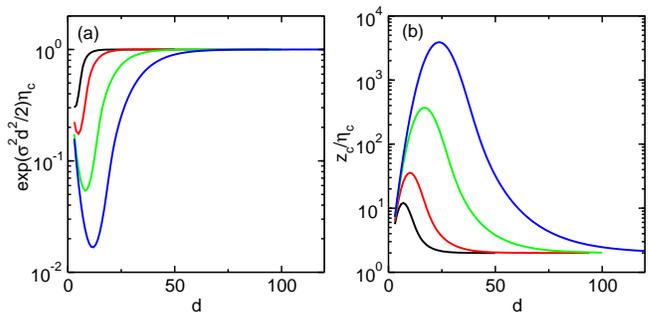}
\caption{(Color online) (a) Dimensional dependence of the percolation threshold $\eta_c$ in units of the asymptotic 
value $\exp(-\sigma^2 d^2/2)$ for a lognormal distribution of the radii obtained from a numerical solution of Eq.~\eqref{log2}; 
$\sigma=0.25$, $0.3$, $0.4$, and $0.5$ from the lowermost to the uppermost curves.
(b) Critical average connectivity per particle $z_c$ for $\sigma=0.25$, $0.3$, $0.4$, and $0.5$ from the uppermost to the lowermost curves. 
All curves tend to $z_c/\eta_c\rightarrow 2$ as $d\rightarrow \infty$.}\label{fig5}
\end{center}
\end{figure}

As evidenced in Eq.~\eqref{log7}, the percolation threshold for infinite dimensions is no longer universal, as it depends on the
parameter $\sigma$ of the log-normal distribution. Interestingly, from Eq.~\eqref{log7} we also see that $\eta_c$ can be
smaller than the critical threshold of monodisperse spheres ($\eta_c=2^{-d}$), contrary to what is expected in finite 
dimensions \cite{Meester1994}. We note that a critical threshold smaller than the monodisperse sphere limit in large dimensions
has been found also for the case of radii distributions with $d$-dependent weights \cite{Gouere2013,Gouere2014}.

To verify the accuracy of Eq.~\eqref{log7}, we compare it with the threshold obtained by solving numerically Eq.~\eqref{log2}. 
As $d$ increases, the asymptotic limit $\eta_c=e^{-\frac{1}{2}\sigma^2d^2}$ is reached more rapidly when $\sigma$ is larger, 
as shown in Fig.~\ref{fig4}(a). From inspection of Eq.~\eqref{log3} we see that this behavior is due to the competition between 
$e^{\sigma^2k(k-d)}$ and the maximum value $\sim 2^d$ of the binomial coefficient at $k= d/2$: the latter is suppressed by 
the exponential function when $d>4\ln(2)/\sigma^2$. From numerical calculation of $z_c$, shown in Fig.~\ref{fig5}(b) for the same
$\sigma$ values of Fig.~\ref{fig5}(a), we see that also the asymptotic formula for $z_c$, Eq.~\eqref{log8}, is verified.

\section{Lower bound on the percolation threshold}
\label{bound}

Having established that Eqs.~\eqref{T2} and \eqref{S3} give asymptotic limits of the critical threshold $\eta_c$ as $d\rightarrow\infty$, 
we show now that the same equations provide also a lower bound on $\eta_c$ for any dimensionality. Toward that end, we take the
pair-connectedness function $P_{ij}(\mathbf{r})$ considered in Sec.~\ref{OZ}, and we extend to the polydisperse sphere case the inequality
formulated in Ref.~\cite{Given1990}:
\begin{equation}
\label{given1}
P_{ij}(\mathbf{r})\leq f_{ij}(r)+\rho\sum_l x_l\int d\mathbf{r}'f_{il}(\vert\mathbf{r}-\mathbf{r}'\vert)P_{lj}(\mathbf{r}'),
\end{equation}
where $f_{ij}(r)$ is the connectedness function given in Eq.~\eqref{connect}. The above expression applies to any dimensionality,
and following Ref.~\cite{Torquato2012}, where Eq.~\eqref{given1} has been used for the monodisperse sphere case,  it enable us to find 
a lower bound on the percolation threshold. To see this, we take the volume integral of Eq.~\eqref{given1},
\begin{equation}
\label{given2}
P_{ij}\leq V_\textrm{ex}^{ij}+\rho\sum_l x_l V_\textrm{ex}^{il}P_{lj},
\end{equation}
where $V_\textrm{ex}^{ij}=\int d\mathbf{r}f_{ij}(r)=\Omega_d(R_i+R_j)^d$, and we use Eqs.~\eqref{OZ3} to find:
\begin{equation}
\label{given3}
\widetilde{T}_i\leq 1+\rho\sum_j x_j V_\textrm{ex}^{ij}\widetilde{T}_j=1+\sum_j z_{ij}\widetilde{T}_j,
\end{equation}
which together with Eq.~\eqref{OZ4a} gives an upper bound for the mean cluster size:
\begin{equation}
\label{given4}
S=\sum_i x_i\widetilde{T}_i\leq\sum_i x_i T_i,
\end{equation}
where $T_i$ is the solution of Eq.~\eqref{T2}. From the inequality of Eq.~\eqref{given4}, we see that the value of $\eta$ such that
$\sum_i x_i T_i$ diverges identifies a lower bound on the percolation threshold for any $d$. The solid lines plotted in Figs.~\ref{fig3}(a)-\ref{fig5}(a) 
represent thus lower bounds on $\eta_c$ for the different radii distribution functions considered in this work. As $d$ increases, 
these lower bounds tend asymptotically to the infinite dimensional limit $2^{-d}$ for bounded radii distributions and to
$\exp(-\sigma^2 d^2/2)$ for lognormal radii distributions.
Finally, we note that Eq.~\eqref{given4} implies also that the values of $z_c$ shown in Figs. ~\ref{fig3}(b)-\ref{fig5}(b) are lower bounds on the 
critical connectivity for any dimensionality.

\section{Summary and discussion}
\label{concl}
We have considered random dispersions of penetrable $d$-dimensional spheres with distributed radii in terms of weighted random geometric graphs,
where nodes represent sphere centers and edges connect nodes of overlapping spheres with probability weighted by the sphere radii.
For bounded distribution of the radii, we have shown that closed loops of connected spheres can be neglected in the limit $d\rightarrow\infty$
and that graph components have thus tree-like structure. Analysis of the mean cluster size reveals that the asymptotic percolation threshold
is universal and coincides with the threshold $\eta_c=2^{-d}$ found for the case of monodisperse spheres in high dimensions. 
This result confirms and extends a previous finding on the percolation of $d\rightarrow\infty$ spheres with two different radii \cite{Gouere2013,Gouere2014}.
Furthermore, we show that the asymptotic critical connectivity per particle $z_c$, though dependent on the shape of the radii 
distribution function, is less than unity and approaches $z_c\rightarrow 1$ for spheres of identical radii. 

We have also studied critical connectivity for spheres with radii distributed according to a $d$-independent lognormal function, which is a treatable 
example of unbounded distribution. Assuming that clusters have a tree-like structure, we find that the percolation threshold $\eta_c$ depends on
the shape of the log-normal distribution and, interestingly, that $\eta_c$ for $d\rightarrow \infty$ can be smaller that the threshold 
for monodisperse spheres, in contrast to what is expected at finite dimensions \cite{Meester1994}.  

Before concluding, let us speculate on the percolation threshold in homogeneous fluids of polydisperse spheres with impenetrable cores 
(cherry-pit model \cite{TorquatoBook}).
In finite dimensions, correlations between the cores preclude writing the multi-degree distribution as a product of Poisson distributions, as done
in Sec.~\ref{multi}, because the $N$-particle distribution function $g_N(\mathbf{r}_1,\mathbf{r}_2,\ldots ,\mathbf{r}_N)$ depends
on the relative positions of the core centers \cite{HansenMcDonald}. However, in the limit of infinite dimensions and for small densities, 
$g_N(\mathbf{r}_1,\mathbf{r}_2,\ldots ,\mathbf{r}_N)$ asymptotically factorizes into a product of $\theta$-functions that are unity for 
pair distances beyond the hard-core diameter \cite{Torquato2006}.
The multi-degree distribution for $d\rightarrow\infty$ can thus still be written as a product of Poisson distributions, with the average number of
contacts unaltered by the presence of the hard-cores if the penetrable shells are non-vanishing. 
With the same reasoning, closed loops are expected to be negligible and graphs
are still dominated by tree-like components.  For non-zero penetrable shells, therefore we expect the same asymptotic results for $\eta_c$
as those obtained for the case of penetrable hyperspheres.
 
\acknowledgements
I am grateful to Avik P. Chatterjee, J.-B. Gou\'er\'e, and S. Torquato for useful comments and suggestions.

\appendix
\section{Irrelevance of $\langle c_d\rangle^{(n)}$ for $d\rightarrow \infty$}
\label{appa}

In this appendix, we show that when the radii distribution is independent of $d$ and bounded [that is, when $\rho(R)=0$ for
any $R>R_M$, with $R_M<\infty$], the $n$-cycle coefficient for polydisperse spheres, defined in Eqs.~\eqref{ncycle4}-\eqref{Vexn}, 
vanishes for $d\rightarrow\infty$. 

Since $R_M$ is the maximum radius of the distribution, the connectedness functions in the integrand of Eq.~\eqref{ncycle5} 
are such that $f_{ij}(r)\leq f(r)=\theta(2R_M-r)$ for any $i$ and $j$. We can thus write:
\begin{equation}
\label{app1}
\mathcal{C}_{i_1,\ldots, i_n}^{(n)}\leq \int dr^{(n)}f(\vert\mathbf{r}_1-\mathbf{r}_2\vert)f(\vert\mathbf{r}_2-\mathbf{r}_3\vert)
\cdots f(\vert\mathbf{r}_n-\mathbf{r}_1\vert),
\end{equation}
which, when substituted in Eq.~\eqref{ncycle4}, gives:
\begin{equation}
\label{app2}
\langle c_d\rangle^{(n)}\leq c^{(n)}_d\frac{VV_\textrm{ex}^{n-1}}{\langle \mathcal{V}_{i_1,\ldots ,i_n}^{(n)}\rangle_{i_1,\ldots ,i_n}},
\end{equation} 
where $V_\textrm{ex}=2^d\Omega_d R_M^d$, and $c^{(n)}_d$ is the $n$-cycle coefficient for identical radii given in 
Eq.~\eqref{ncycle3}. Next, we perform the integrations over $\mathbf{r}_1,\ldots,\mathbf{r}_n$ in Eq.~\eqref{Vexn} to find:
\begin{align}
\label{app3}
\mathcal{V}_{i_1,\ldots ,i_n}^{(n)}&=V\prod_{j=1}^{n-1}V_\textrm{ex}^{i_j,i_{j+1}}=V\Omega_d^{n-1}\prod_{j=1}^{n-1}(R_{i_j}+R_{i_{j+1}})^d\nonumber \\
&=V\Omega_d^{n-1}\prod_{j=1}^{n-1}\left[\sum_{k_j=0}^d\binom{d}{k_j}R_{i_j}^{k_j}R_{i_{j+1}}^{d-k_j}\right],
\end{align} 
where in the last equality we have expanded the binomial powers. In performing the average over $R_{i_1},\ldots,R_{i_n}$, we must
group the contributions with equal radius variables and average them independently of the other radii. Denoting a general
$m$-th moment as $\langle R^m\rangle_R$, we obtain:
\begin{align}
\label{app4}
\langle \mathcal{V}_{i_1,\ldots ,i_n}^{(n)}\rangle_{i_1,\ldots ,i_n}&
=V\Omega_d^{n-1}\sum_{k_1=0}^d\binom{d}{k_1}\cdots\sum_{k_{n-1}=0}^d\binom{d}{k_{n-1}}\nonumber \\
&\times\langle R^{k_1}\rangle_R\langle R^{d-k_1+k_2}\rangle_R\langle R^{d-k_2+k_3}\rangle_R\cdots\nonumber \\
&\cdots\langle R^{d-k_{n-2}+k_{n-1}}\rangle_R\langle R^{d-k_{n-1}}\rangle_R.
\end{align}
Following Sec.~\ref{continuous}, we approximate for large $d$ the binomial coefficients by Gaussian functions centered at $d/2$, and we
replace the sums by integrals, so that for $d\rightarrow\infty$,  $\sum_{k_j=0}^d\binom{d}{k_j}\rightarrow 2^d\int_0 ^2 ds_j \delta(s_j-1)$,
where $s_j=2k_j/d$ and $j=1,\ldots,n-1$. Equation~\eqref{app4} reduces in this way to:
\begin{equation}
\label{app5}
\langle \mathcal{V}_{i_1,\ldots ,i_n}^{(n)}\rangle_{i_1,\ldots ,i_n}\rightarrow V\Omega_d^{n-1}2^{(n-1)d}\langle R^{d/2}\rangle_R^2\langle R^d\rangle_R^{n-2},
\end{equation}
so that Eq.~\eqref{app2} becomes:
\begin{equation}
\label{app6}
\langle c_d\rangle^{(n)}\leq c^{(n)}_d\chi_d^{(n)},
\end{equation}
where
\begin{equation}
\label{app7}
\chi_d^{(n)}=\frac{R_M^{(n-1)d}}{\langle R^{d/2}\rangle_R^2\langle R^d\rangle_R^{n-2}}.
\end{equation}
For continuous radii distributions, we assume that $\rho(R)\propto (R_M-R)^{\alpha-1}$ with $\alpha>0$ for $R\rightarrow R_M$, as done in Sec.~\ref{continuous}.
Using Eq.~\eqref{f2} we thus find $\chi_d^{(n)}\propto d^{n\alpha}$. For discrete distributions as in Eq.~\eqref{fin1}, it is straightforward
to show from Eq.~\eqref{app7} that $\chi_d^{(n)}\rightarrow 1/(x_M)^n$ for large $d$. We have thus arrived at the result that $\chi_d^{(n)}$ increases with
$d$ at most as a power-law, leading to $\langle c_d\rangle^{(n)}\rightarrow 0$ as $d\rightarrow\infty$, due to the exponential vanishing of $c^{(n)}_d$.

\end{document}